\documentclass[useAMS,usenatbib]{mn2e}

\usepackage{epsfig,graphicx,caption} 
\usepackage{verbatim}
\graphicspath{{./}}

\title[A parametric physical model for the ICM]{A
  parametric physical model for the intracluster medium and its use in
  joint SZ/X-ray analyses of galaxy clusters} \author[J.~R. Allison et
al.]{J.~R. Allison$^{1}$\thanks{E-mail: jra@astro.ox.ac.uk},
  A.~C. Taylor$^{1}$, M.~E. Jones$^{1}$, S. Rawlings$^{1}$, and
  S.~T. Kay$^{2}$\\$^{1}$Oxford Astrophysics, Denys Wilkinson
  Building, Keble Road, Oxford OX1 3RH, U.K.\\$^{2}$Jodrell Bank
  Centre for Astrophysics, School of Physics \& Astronomy, The
  University of Manchester, Manchester M13 9PL, U.K.}

\begin{document}

\date{}

\pagerange{\pageref{firstpage}--\pageref{lastpage}} \pubyear{2010}

\maketitle

\label{firstpage}

\begin{abstract}
  We present a parameterized model of the intra-cluster medium that is
  suitable for jointly analysing pointed observations of the
  Sunyaev-Zel'dovich (SZ) effect and X-ray emission in galaxy
  clusters. The model is based on assumptions of hydrostatic
  equilibrium, the Navarro, Frenk and White (NFW) model for the dark
  matter, and a softened power law profile for the gas entropy. We
  test this entropy-based model against high and low signal-to-noise
  mock observations of a relaxed and recently-merged cluster from
  $N$-body/hydrodynamic simulations, using Bayesian hyper-parameters
  to optimise the relative statistical weighting of the mock SZ and
  X-ray data. We find that it accurately reproduces both the global
  values of the cluster temperature, total mass and gas mass fraction
  ($f_\rmn{gas}$), as well as the radial dependencies of these
  quantities outside of the core ($r > 100\,\rmn{kpc}$). For reference
  we also provide a comparison with results from the single isothermal
  $\beta$ model. We confirm previous results that the single
  isothermal $\beta$ model can result in significant biases in derived
  cluster properties.
\end{abstract}


\section{Introduction}\label{Introduction}

The Sunyaev-Zel'dovich effect is a distortion of the Cosmic Microwave
Background (CMB) spectrum by inverse Compton scattering off a hot
population of electrons, such as in the intra-cluster medium (ICM) of
a galaxy cluster \citep{Sunyaev:1970, Birkinshaw:1999}. The SZ effect
is observable as an anisotropy in the CMB and has a surface brightness
that is proportional to the line-of-sight integral of the gas pressure
of the cluster, independent of redshift. Since the surface brightness
is independent of redshift, SZ observations provide a powerful tool to
probe the properties of the largest virialised structures in the
Universe. Over the past decade there has been considerable
observational data gathered on the ICM properties of clusters via the
SZ effect \citep[e.g.][]{Grego:2001, LaRoque:2006} and we are now
entering an era of deep cluster surveys with specifically designed SZ
telescopes, including the South Pole Telescope \citep[SPT,
][]{Ruhl:2004}, the Sunyaev-Zel'dovich Array \citep[SZA,
][]{Muchovej:2007}, the Atacama Cosmology Telescope \citep[ACT,
][]{Kosowsky:2003} and the Arcminute Microkelvin Imager \citep[AMI,
][]{Zwart:2009}. These surveys will make use of the redshift
independence of the SZ surface brightness to map the distribution and
evolution of large scale structure throughout the Universe.

In order to interpret the data from SZ surveys we require
well-calibrated scaling relations between the SZ signal and physical
properties and a good physical understanding of the intrinsic scatter
therein. Pointed SZ observations with interferometer and bolometer
arrays coupled with X-ray imaging and spectra allow us to probe the
intra-cluster medium, while weak and strong gravitational lensing
observations directly measure the projected gravitational potential
and hence total mass of the cluster. Recently, there has been
concerted effort in developing physically motivated analytic models
that account for the observed non-isothermality in cluster gas [see
e.g. \citealt{Komatsu:2001, Ostriker:2005, Atrio-Barandela:2008,
  Bulbul:2010} (polytropic equation of state),
\citealt{Pointecouteau:2004, Vikhlinin:2006, Mahdavi:2007} (component
separation and modified $\beta$ models), and \citealt{Nagai:2007,
  Mroczkowski:2009, Arnaud:2010} (explicit pressure parametrization)].

We present a model in this work that avoids using an explicit
temperature and electron density parametrization that may lead to
non-physical cluster properties on large scales. We instead choose to
base our ICM model on a simple parametrization of the cluster
entropy, consistent with X-ray observations and cluster theory of
spherical shock accretion and cooling. We use the NFW parametrization
\citep{Navarro:1997} for the total mass distribution and assume
hydrostatic equilibrium. We outline our model with reference to
pointed SZ observations with the Cosmic Background Imager 2 experiment
(CBI2) located at the Chajnantor observatory in the Atacama Desert,
Chile. This was an interferometer operating at 26--36\,GHz, with
10\,$\times$\,1\,GHz channels, and thirteen 1.4\,m antennas
\citep[][Taylor et al. in prep.]{Padin:2002}. CBI2 has a 28\,arcmin
field of view and 6\,arcmin resolution so that at moderate redshifts
($z \sim 0.3$) it can observe out to the largest radii of galaxy
clusters. In future work we will present the application of the
procedure outlined here to observations of a number of massive galaxy
clusters with CBI2.

We adopt a $\Lambda$CDM cosmology with $\Omega_\rmn{M} = 0.3$,
$\Omega_\lambda = 0.7$ and $H_0 = 70$\,km\,s$^{-1}$\,Mpc$^{-1}$
throughout this paper unless otherwise stated.

\section{Modelling the Intra-Cluster Medium}\label{modelling_the_ICM}

\subsection{Current Models of the intra-cluster medium}

In the past most of the joint analysis of SZ and X-ray data from
observations of galaxy clusters made use of the isothermal $\beta$
model prescription for modelling the ICM
\citep{Cavaliere:1976,Cavaliere:1978}. This model is typically used to
fit regions of the cluster within $r_{2500}$, the radius at which the
mean enclosed density is 2500 times the universal critical density at
the redshift of the cluster. At relatively small radii the isothermal
$\beta$ model is found to be accurate in reproducing the average
observable and physical properties of many clusters \citep[see
e.g.][]{Grego:2000, Reese:2002, LaRoque:2006}. However, deep X-ray
spectral observations of nearby cluster samples using \emph{Chandra}
\cite[see e.g.][]{Vikhlinin:2006} and \emph{XMM-Newton} \cite[see
e.g.][]{Pratt:2007} show that at larger radii the ICM temperature
declines with radius. Hence for low-resolution pointed SZ observations
that are sensitive to the outskirts of the cluster gas, there is a
clear requirement for a more physically motivated non-isothermal model
that is simple enough to be constrained by the data.

There have been a number of recent approaches amongst SZ and X-ray
observers to relax the assumption of isothermality in clusters and
model the ICM in a physical way. For \emph{Chandra} observations of
nearby relaxed clusters, \cite{Vikhlinin:2006} developed a modified
$\beta$ model of the electron density to fit the observed cores and
steeper outer profiles of relaxed clusters. In that work the
temperature is parameterized by a broken power law at large radii and
declines in the central cooling region as given by
\cite{Allen:2001}. The \scriptsize JACO \normalsize software pipeline
developed by \cite{Mahdavi:2007} uses separate parametrization of the
gas, dark matter and stellar components, assuming hydrostatic
equilibrium, to jointly fit to high significance X-ray spectra from
\emph{Chandra} and \emph{XMM-Newton}, weak lensing shear from
Canada-France-Hawaii telescope, optical data on the brightest cluster
galaxy from the Hubble space telescope and the SZ effect from the
Cosmic Background Imager. For sufficiently resolved SZ maps, the
temperature and mass profiles can be reconstructed by de-projection
techniques, for example see \cite{Ameglio:2007,Ameglio:2009} for a
comparison of the method to simulations and \cite{Nord:2009} for
direct application to SZ and X-ray data. However with many SZ
experiments the data are generally not very well resolved, and so
combined analysis with X-ray surface brightness data requires the
application of a parameterized model that reproduces the observed
temperatures and masses from X-ray observations and
simulations. \cite{Mroczkowski:2009} successfully use an NFW
parameterized pressure profile developed by \cite{Nagai:2007} to fit
to data from the Sunyaev-Zel'dovich Array (SZA), and then combine with
X-ray surface brightness to reproduce temperature profiles out to
large radii. This model has also been adopted by \cite{Arnaud:2010} in
order derive a universal pressure profile from \emph{XMM-Newton}
observations of the REXCESS cluster sample. Recently
\cite{Bulbul:2010} have developed an analytical cluster model based on
assuming a polytropic relationship between the gas density and
temperature, and parameterise the total mass using a generalised NFW
model. In order to account for cool-core behaviour the resulting
radial expressions are modified by the same core taper used by
\cite{Vikhlinin:2006}.

The CBI2 array has relatively low spatial resolution, and a large
field-of-view, and hence a simple parameterized model is required that
is accurate over the bulk of the volume of the cluster. We could
choose any function of gas density and temperature as the basis for
our gas model, e.g. pressure ($\propto Tn_\rmn{e}$) or entropy
($\propto Tn_\rmn{e}^{-2/3}$). For example \cite{Nagai:2007} choose to
use a four-parameter model for the cluster pressure profile. We choose
to parameterize the ICM based on the entropy because there is evidence
from both observations and simulations that it can be modelled as a
simple power-law over most of the cluster volume (see
Section\,\ref{section:entropy_model} below). We combine this with an
NFW parametrization of the dark matter halo component. By modelling
the gas in this way, the underlying physics of cluster gas in
hydrostatic equilibrium is encoded into the model from the outset, and
is simple enough to be constrained by X-ray surface brightness and
low-resolution SZ data. This also avoids ad-hoc parametrization of
the electron density and temperature and therefore removes the
possibility of introducing unphysical solutions for the total
mass. The following section describes in detail the physical basis and
parametrization of this model.

\subsection{An Entropy-based model}\label{section:entropy_model}

\subsubsection{The Dark Mass Halo}

We construct a parametric model that is physically consistent and is
therefore a reasonable description of the ICM on the angular scales to
which CBI2 is sensitive. A suitable parametrization of the dark
matter content is the hierarchical clustering NFW profile
\citep{Navarro:1995}, where the dark matter (DM) density as a function
of radius is given by
\begin{equation}
  \rho_\rmn{DM}(r)= {\delta_\rmn{c}\rho_\rmn{crit}
    \over(r/r_\rmn{s})(1 + r/r_\rmn{s})^2},
\end{equation}
where $\rho_{\rmn{crit}}$ is the universal critical density for
closure (at the redshift of the cluster), $r_{\rmn{s}}$ is the scale
radius and $\delta_{\rmn{c}}$ is the characteristic density
contrast. The DM virial radius $r_\rmn{DM}$ is defined in this model
as the radius of a sphere at which the mean interior DM density is
equal to $200\rho_{\rmn{crit}}$. Note that the true virial radius of
the cluster is slightly larger ($\sim 10\,\rmn{per cent}$) since one
must also take into account the contribution from the gas when
calculating the total mass. It is related to the scale radius by
\begin{equation}
  r_{\rmn{s}} = {r_\rmn{DM}\over c_\rmn{DM}},
\end{equation}
where $c_\rmn{DM}$ is known as the DM concentration parameter. This
is related to the DM density contrast \citep{Navarro:1996} by
\begin{equation}
  \delta_{\rmn{c}} = {200\over3}{c_\rmn{DM}^3\over[\ln(1 + c_\rmn{DM}) -
    c_\rmn{DM}/(1 + c_\rmn{DM})]}.
\end{equation}
The DM density becomes increasingly large at small radii, and is
unbounded at the centre of the cluster. However the enclosed DM mass
is calculated by integrating this density over volume and this
converges at the origin. The enclosed DM mass is given by the
following expression
\begin{equation}
  M_\rmn{DM}(<r) =
  4\pi\delta_{\rmn{c}}\rho_{\rmn{crit}}r_\rmn{s}^3\left[\ln\left({r_\rmn{s}
        + r\over r_\rmn{s}}\right) - \left({r\over r_\rmn{s} +
        r}\right)\right].
\end{equation}
Therefore the DM mass distribution in the cluster is defined by
the two parameters $c_\rmn{DM}$ and $r_\rmn{DM}$.

\subsubsection{The Entropy}

In order to derive the electron pressure, and hence the thermal SZ
decrement, a physical quantity is required that describes the gaseous
properties in a similar way for all clusters. The cluster entropy is
expected to behave in a self-similar way for reasonably massive
clusters, differing primarily in the cluster cores due to the
injection of non-gravitational thermal energy. The electron pressure
and entropy observable \citep{Ponman:1999, Lloyd-Davies:2000} are
given by
\begin{eqnarray}
  P_\rmn{e} &=& S_\rmn{e}\,n_\rmn{e}^{\gamma} \\ 
  S_\rmn{e} &=& {T_\rmn{e}\over n_\rmn{e}^{\gamma-1}}, \label{equation:entropy_def}
\end{eqnarray}
where $P_\rmn{e}$, $S_\rmn{e}$, $n_\rmn{e}$ and $T_\rmn{e}$ are the
electron pressure, entropy, density and temperature (in keV units)
respectively. The parameter $\gamma$ is the adiabatic ratio of heat
capacities and for an ideal monatomic gas is equal to 5/3. The
observable quantity $S_\rmn{e}$ is related to the true thermodynamic
entropy per gas particle $S_\rmn{th}$ by the following equation
\begin{equation}
S_\rmn{th} = A\ln(S_\rmn{e}) + B,
\end{equation}
where $A$ and $B$ are both functions of the fundamental and ideal gas
constants. At small radii \citep[$r \la 0.1r_{200}$,][]{Voit:2005} the
entropy profiles of clusters are expected to have a low-entropy core
region as a result of radiative cooling of the gas. However excess
entropy may also exist in this region due to additional heating
processes from central AGN and star formation. At the opposite end of
the radial scale the accretion of matter at the interface between the
cluster gas and the external medium generates an expanding
shock-front, with an ever-increasing virial mass and hence
infall-velocity. It is therefore expected that the entropy profile
increases as a function of radius towards the edge of the cluster.
Simulations of non-radiative clusters predict that the entropy profile
should tend to a power law, where $S_\rmn{e}(r) \propto r^{1.1}$
\citep[e.g.][]{Tozzi:2001, Kay:2004a, Voit:2005, Mitchell:2009}. X-ray
observations to date are consistent with this predicted behaviour
\citep[see e.g.][]{Ponman:2003, Piffaretti:2005, Donahue:2006,
  Pratt:2006, Morandi:2007, Zhang:2008, Pratt:2010}, and a recent
study of the high resolution \emph{Chandra} Data Archive by
\cite{Cavagnolo:2009} has provided evidence of this expected behaviour
in a collection of 239 clusters. Therefore, for the purposes of
providing a physically motivated gas model, we choose the
beta-model-like parametrization for the entropy profile
\begin{equation}
S_\rmn{e}(r) = S_\rmn{e0}\left(1 + \left(r\over
r_{\rmn{core}}\right)^2\right)^{\alpha},
\end{equation}
where $S_\rmn{e0}$ is the central entropy value, $r_{\rmn{core}}$ is a
scale radius, and $\alpha$ is a parameter describing the behaviour at
radii much larger than $r_\rmn{core}$. In terms of the central
electron density and temperature, $S_\rmn{e0} =
T_\rmn{e0}/n_{\rmn{e}0}^{2/3}$. Given the findings of the simulation
and observational work outlined above, one would expect the entropy
profile to be dominated by non-gravitational processes at radii
smaller than $r_\rmn{core} \sim 0.1r_{200}$, and at larger radii
follow a power-law $S_\rmn{e} \propto r^{2\alpha}$, where $\alpha \sim
0.55$.

We note that recent X-ray observations of the ICM using \emph{Suzaku}
give evidence for a deviation in the entropy behaviour from the
$r^{1.1}$ power-law to flatter profiles at large radii
\citep[e.g.][]{George:2009, Bautz:2009, Hoshino:2010,
  Kawaharada:2010}. Interpretations of these results include
significant deviation from hydrostatic equilibrium and the effects of
low thermalisation between electrons and ions in the cluster
outskirts, especially where the edge of the cluster gas meets
low-density regions \citep[see][and references
therein]{Kawaharada:2010}. The entropy-based model presented here can
easily be adapted to account for this observed behaviour at large
radii by the introduction of additional parameters. For the purposes
of this work we do not adapt the model for the results of the
\emph{Suzaku} observations, however in future work on CBI2 data we
will consider the entropy behaviour at the cluster outskirts.

\subsubsection{The Pressure}

If we assume that the gas particles of the intra-cluster medium are in
local thermodynamic equilibrium, then the condition of hydrostatic
equilibrium relates the total enclosed mass ($M$) to the electron
density and partial pressure by
\begin{equation}\label{equation:HE}
{dP_\rmn{e}\over \rmn{d}r} = -{GM\mu m_\rmn{p}n_\rmn{e}(r)\over r^2},
\end{equation}
where $\mu$ is the mean molecular mass per gas particle, which has a
value of 0.6 for a fully ionised plasma with cosmic hydrogen and
helium abundance ratios. From the expression given in
Equation\,\ref{equation:entropy_def}, and assuming the ideal gas law
holds for the electron pressure (where $P_\rmn{e} =
n_\rmn{e}T_\rmn{e}$), the electron density as a function of pressure
and entropy can be expressed as
\begin{equation}
n_\rmn{e} = (P_\rmn{e}/S_\rmn{e})^{3/5}.
\end{equation}
The pressure as a function of radius can then be derived by
substituting the above expression for $n_\rmn{e}$ into
Equation\,\ref{equation:HE} and rearranging to give
\begin{equation}\label{equation:pressure}
  P_\rmn{e}(r) = \left[P_\rmn{e0}^{2/5} - {2\over5}G\mu m_\rmn{p}\int_0^{r}{Mdr'\over
      S_\rmn{e}^{3/5}r'^2}\right]^{5/2},
\end{equation}
where the central pressure, $P_\rmn{e0}$, is equal to the product
of the central density and temperature. The integral in the above
expression can be solved numerically so that the electron pressure is
calculated as a function of radius. The total mass is calculated by
summing the enclosed DM and baryonic masses (note that the fraction of
mass contained in stellar material is considered insignificant) at
each radius out from the centre of the cluster. A value of 1.16 is
assumed for the ratio of baryons to electrons, in order to calculate
the baryonic mass. The pressure is then integrated in shells from the
cluster centre to the required radius.

\subsection{SZ and X-ray Observables}

The SZ signal is a measure of the integrated line-of-sight electron
pressure, and when integrated over the whole cluster is a measure of
the total thermal energy. The observed change in CMB temperature as a
result of the SZ effect is given by
\begin{equation} 
  {\Delta T_\rmn{SZ} \over T_\rmn{CMB}} = f(x)y,
\end{equation}
where $x$ is a dimensionless frequency equal to
$h\nu/k_\rmn{B}T_\rmn{CMB}$ and $y$ is the comptonisation
parameter, which is equal to the integrated optical depth multiplied
by the fractional gain in energy per scattering. Hence $y$ is
proportional to the integral of $n_\rmn{e}T_\rmn{e}$ along the
line of sight
\begin{equation}
  y = \int n_\rmn{e}{k_\rmn{B}T_\rmn{e}\over
    m_\rmn{e}c^2}\sigma_\rmn{T}\rmn{d}l,
\end{equation}
The frequency dependency $f(x)$ also has a dependence on the electron
temperature $T_\rmn{e}$, given by
\begin{equation}
  f(x, T_\rmn{e}) = \left(x{\rmn{e}^{x}+1\over \rmn{e}^{x}-1} - 4\right)(1 +
  \Delta(x, T_\rmn{e})),
\end{equation}
where the relativistic correction $\Delta(x, T_\rmn{e})$ is derived
from higher-order corrections to the Kompaneets equation
\citep{Challinor:1998, Itoh:1998}.

In the general case of modelling a non-isothermal cluster the product
of pressure and $f(x, T_\rmn{e})$ is integrated along the
line-of-sight. Replacing $n_\rmn{e}$ and $T_\rmn{e}$ by the
radial dependence of the electron pressure, the expression for the
change in the CMB temperature is given by
\begin{equation}\label{equation:sz_temp}
  \Delta T_\rmn{SZ} = T_\rmn{CMB}\int f(x, T_\rmn{e}) {P_\rmn{e}\sigma_\rmn{T}\over
    m_\rmn{e}c^2}\rmn{d}l,
\end{equation}
which is then solved numerically.

The integral of pressure over the volume of the cluster is a measure
of the total thermal energy and is therefore a useful quantity in
characterising the overall thermal SZ effect. The integral of $y$ over
the solid angle of the cluster is an observable quantity that is often
used for this purpose,
\begin{equation}
Y = \int{y(\Omega)\rmn{d}\Omega},
\end{equation}
where $Y$ is calculated within a given projected radius. However, for
the purposes of this work, we choose to use an intrinsic measure of
$Y$ that is independent of the distance to the cluster, and hence will
allow comparison of clusters at different redshifts. We therefore
assume spherical symmetry and use the following definition
\begin{equation}
Y\,D_\rmn{A}^{2} = 4\pi\int{{P_\rmn{e}\sigma_\rmn{T}\over m_\rmn{e}c^{2}}r^{2}\rmn{d}r},
\end{equation}
where $D_\rmn{A}$ is the angular diameter distance to the cluster and
$Y\,D_\rmn{A}^{2}$ is calculated over the cluster volume, enclosed by
a physical radius typically equal to the virial radius.
  
The observed continuum X-ray surface brightness from galaxy clusters
is caused primarily by bremsstrahlung radiation from scattered
electrons in the hot ICM, with some emission also from high energy
transition lines. The continuum X-ray emission is proportional to the
projected square of the electron density, multiplied by the emissivity
along the line-of-sight. If all of the emission were completely
dominated by bremsstrahlung radiation then the emissivity would go as
the square root of the temperature. While in practise this is not the
complete case, in general the emissivity is a weak function of the
temperature. The observed X-ray surface brightness per unit energy
$\rmn{d}E$ is given by the following expression
\begin{equation}
  {\rmn{d}S_\rmn{X}\over \rmn{d}E} = {1\over 4\pi(1+z)^{3}}\int{n_\rmn{e}^{2}\tilde{\Lambda}(E',T_\rmn{e})\rmn{d}l},
\end{equation}
where $z$ is the cluster redshift and $\tilde{\Lambda}(E',T_\rmn{e})$
is the X-ray spectral emissivity as a function of temperature and
emitted energy, $E' = E(1+z)$. The surface brightness for a particular
observer-frame energy band is then given by the equation
\begin{equation}\label{equation:xray_sb}
  S_\rmn{X}  = {1\over 4\pi(1+z)^{4}}\int{n_\rmn{e}^{2}\Lambda(T_\rmn{e})\rmn{d}l},
\end{equation}
where
\begin{equation}
\Lambda(T_\rmn{e}) = \int{\tilde{\Lambda}(E',T_\rmn{e})\rmn{d}E}.
\end{equation}
In this work the emissivity values for every radial position in the
cluster volume are interpolated from a look-up table of values, as a
function of temperature and redshift, based upon the Raymond-Smith
plasma model \citep{Raymond:1977} for an observer-frame energy range
0.5-2\,keV and a metallicity of $Z = 0.3\,Z_{\odot}$.

Combined observations of the SZ and X-ray surface brightness constrain
the pressure and electron density of the ICM and hence provide
constraints on all of the model parameters. The physical properties of
the cluster, including the total mass and electron temperature, as a
function of radius are constrained by the 6 parameters defined by this
model: $c_\rmn{DM}$, $r_\rmn{DM}$, $n_\rmn{e0}$, $T_{e0}$,
$r_\rmn{core}$ and $\alpha$.

\section{Model Fitting}\label{section:model_fitting}

\subsection{Bayesian inference of the parameter values}

We fit parameterized models to the SZ and X-ray data and obtain
probability distributions for the derived physical quantities. The
joint probability distribution of a set of model parameters
($\bmath{\theta}$), given the data ($\bmath{d}$) and the model
($\mathcal{M}_\rmn{j}$), can be calculated from Bayes' Theorem,
\begin{equation}\label{equation:bayes_theorem}
  \rmn{Pr}(\bmath{\theta}|\bmath{d},\mathcal{M}_\rmn{j}) =
      {\rmn{Pr}(\bmath{d}|\bmath{\theta},\mathcal{M}_\rmn{j})\rmn{Pr}(\bmath{\theta}|\mathcal{M}_\rmn{j})\over
        \rmn{Pr}(\bmath{d}|\mathcal{M}_\rmn{j})}.
\end{equation}
The probability of the data given the set of model parameters
$\rmn{Pr}(\bmath{d}|\bmath{\theta},\mathcal{M}_\rmn{j})$ (the likelihood, $L$)
can be calculated based on assumptions of the distribution of the
error in the data. When the data set is large and therefore
quasi-continuous (such as the noise generated in radio
instrumentation), one can approximate the likelihood by the form given
for Gaussian multivariate data \citep[see e.g.][]{Sivia:2006}
\begin{eqnarray}\label{equation:gaussian_likelihood}
  L & \equiv & \rmn{Pr}(\bmath{d}|\bmath{\theta},\mathcal{M}_\rmn{j}) \nonumber \\
  & = &  {1\over\sqrt{(2\pi)^{N}|\mathbfss{C}|}}\exp{\left(-{(\bmath{d} - \bmath{m})^\rmn{t}\mathbfss{C}^{-1}(\bmath{d} - \bmath{m})\over2}\right)},
\end{eqnarray}
where $N$ is equal to the size of $\bmath{d}$, $\textbfss{C}$ is the
covariance matrix of the data, $|\mathbfss{C}|$ is the determinant of
the covariance matrix and $\bmath{m}$ is the vector of model
data. Since we wish to perform joint fits to $k$ independent data
sets, such that $\bmath{d}$ = $\sum{\bmath{d}_\rmn{k}}$, then the
total likelihood is just given by the product of the individual
likelihoods.

The probability of the parameter values given the model,
$\rmn{Pr}(\bmath{\theta}|\mathcal{M}_\rmn{j})$, is known as the prior
distribution, and encodes the prior information on the parameter
values before fitting to the data. For example, the uncertainty in the
position of the cluster centroid might be known from previous
observations and from the pointing accuracy of the instrument. One
might then wish to apply a Gaussian prior to the model cluster
position based on the level of uncertainty. In this work the prior
distributions for the model parameters are uniform distributions
unless indicated otherwise, and comfortably encompass reasonable
values based on previous work.

The normalisation of the posterior probability distribution is given
by the evidence, which is equal to the probability of the data given
the model. The evidence is a measure of the suitability of the model
for fitting to the data and provides a quantitative measure of
selecting between competing models. The evidence for the data,
assuming a particular model, is calculated by marginalising the
product of the likelihood and prior distributions over the model
parameters. This is given by
\begin{equation}\label{equation:evidence}
  \rmn{Pr}(\bmath{d}|\mathcal{M}_\rmn{j}) = \int{\rmn{Pr}(\bmath{d}|\bmath{\theta},\mathcal{M}_\rmn{j})\rmn{Pr}(\bmath{\theta}|\mathcal{M}_\rmn{j})\rmn{d}\bmath{\theta}},
\end{equation}
which follows from Equation\,\ref{equation:bayes_theorem} and the fact
that the integrated posterior distribution is normalised to
unity. When a model provides a good fit to the data, the likelihood
peak will have a high value, and hence the model will have a high
evidence value associated with it. However if the model is
over-complex then there will be large regions of low likelihood within
the prior volume, thus reducing the evidence value for this model, in
agreement with Occam's razor. This provides a mechanism to choose
between competing models $\mathcal{M}_\rmn{j}$.

\subsection{The SZ Likelihood}

The focus of this work is interferometric SZ data, which are assumed
to have Gaussian distributed errors, and so have a likelihood that is
expressed in the form given by
Equation\,\ref{equation:gaussian_likelihood}. The SZ data consist of
complex visibilities from the output of the correlator, which encode
the Fourier transform of the product of the sky brightness and the
primary beam of the instrument. The visibilities only sample the
Fourier transform at discrete coordinates in the Fourier plane,
defined by the configuration of the instrument antennas during the
observation.

Errors in the SZ data originate from three main sources; the thermal
noise, the intrinsic CMB anisotropy and radio point
sources. The total covariance matrix is thus given by the sum of these
individual components,
\begin{equation}
  \mathbfss{C}_\rmn{SZ} = \mathbfss{C}_\rmn{noise} + \mathbfss{C}_\rmn{CMB} + \mathbfss{C}_\rmn{sources}. 
\end{equation}
The thermal noise component originates from the instrument
electronics, the atmosphere and the ground, and since the noises are
not correlated the covariance matrix for this component is
diagonal. Experimental error in the SZ signal also arises from the
presence of intrinsic fluctuations in the CMB that obscure the
cluster. Since these fluctuations are inherent in the sky brightness,
they are correlated in the visibilities, and so result in a
non-diagonal covariance matrix. The uncertainty due to the CMB is
largest on the shortest CBI2 baselines and dominates the error in the
SZ signal at $l \la 1000 $ \citep[see e.g.][]{Udomprasert:2004}.

The remaining principal source of experimental error is the presence
of bright radio point sources, especially near to the centre of the
field, where there will be less attenuation from the primary beam and
more confusion with components of the SZ signal. This error can be
treated either by identifying the point sources in the long-baseline
data or by simultaneously observing with higher resolution
instruments. If one is confident that source variability is not an
issue, then the source fluxes can be subtracted directly from the data
and any residual flux errors incorporated into the covariance
matrix. Alternatively, contamination due to point sources can be
treated during the model fitting stage by the inclusion of a component
of the covariance matrix corresponding to a source model. Scaling this
component by a large pre-factor is approximately equivalent to masking
out the data at the position of the source. This method is discussed
in detail by \cite{Bond:2000}, \cite{Mason:2003} and
\cite{Sievers:2009}. In the work presented here we consider only
simulated mock data and so the presence of strong radio point sources
are safely ignored. However in future work we will provide a full
description of the point source treatment when presenting actual CBI2
SZ data.

\subsection{The X-ray Likelihood}

The raw X-ray surface brightness data are in the form of discrete
photon counts distributed over a photon counting detector, and hence
will have Poisson distributed errors. However, since we assume that
the cluster model is spherically symmetrical, we can radially bin the
X-ray surface brightness so that the number of pixels contributing to
each bin is typically greater than 100 over the radial range of
interest. Under this assumption the likelihood is then well
approximated by the Gaussian form given by
Equation\,\ref{equation:gaussian_likelihood}. The errors in each
radial bin value are uncorrelated and so the associated covariance is
simply given by a diagonal matrix of the variance in the data.

\subsection{Hyper-parameters and combining the data sets}

Radio interferometric SZ and X-ray surface brightness data typically
have sensitivity to very different angular scales and depend
differently on the physical properties that are being
constrained. Therefore constructing an analytical physically-based
model that provides a satisfactory simultaneous fit to both data sets
is difficult, especially if we are unsure about the accuracy of the
error estimates. It is therefore good practise to weight the
likelihoods for these data sets using two hyper-parameters, which can
then be treated as nuisance parameters and marginalised over. The
normalised modified Gaussian likelihood for the data now has the
following expression
\begin{eqnarray}
  L & \equiv & \rmn{Pr}(\bmath{d}|\bmath{\theta},\epsilon,\mathcal{M}_\rmn{j}) \nonumber \\
  & = & {\sqrt{\epsilon^{N}\over(2\pi)^{N}|\mathbfss{C}|}}\exp{\left(-\epsilon{(\bmath{d} - \bmath{m})^\rmn{t}\mathbfss{C}^{-1}(\bmath{d} - \bmath{m})\over2}\right)},
\end{eqnarray}
where $\epsilon$ is the hyper-parameter. Its assumed that very little
is known about the true value of $\epsilon$, except that it has an
expectation value of unity. In this case a suitable prior for each
hyper-parameter is given by \cite{Hobson:2002}, where
\begin{equation}
\rmn{Pr}(\epsilon) = \exp{(-\epsilon)},
\end{equation}
and $\epsilon$ has a value between 0 and $\infty$. Note that if the
Jeffreys prior were used instead, where $\rmn{Pr}(\epsilon) =
1/\epsilon$, then it would not be possible to normalise the posterior
distribution or calculate the evidence. By calculating the ratio of
the Bayesian evidence for the case where hyper-parameters are used, to
where they are not, it can be inferred whether they should be
included. If this ratio is much larger than unity then there is a
significant relative inconsistency in the error estimates of the two
data sets and their inclusion is warranted.

\subsection{Implementation}

Implementation of the model fitting is performed using \scriptsize
BAYESYS \normalsize\citep{Skilling:2004}, a powerful Bayesian
optimiser that uses the Markov Chain Monte Carlo method. Estimates of
the posterior probability distribution are obtained for the model
parameters by using all of the proposal distribution engines available
to the program. \scriptsize BAYESYS \normalsize provides the modelling
code with parameter samples in the form of a unit hyper-cube from
which the SZ and X-ray surface brightness likelihoods are
calculated. The program initialises with a burn-in phase that evolves
the MCMC chains from sampling just the prior distribution to locating
regions of significant posterior probability. This is performed in a
series of steps known as selective annealing (due to the analogous
relationship with annealing in thermodynamics), during which samples
of parameters are drawn from a modified posterior given by
\begin{equation}
  \rmn{Pr}(\bmath{\theta}|\bmath{d},\mathcal{M}_\rmn{j}) \propto
  \rmn{Pr}(\bmath{d}|\bmath{\theta},\mathcal{M}_\rmn{j})^{\lambda}\rmn{Pr}(\bmath{\theta}|\mathcal{M}_\rmn{j}),
\end{equation}
where the cooling factor ($\lambda$) is slowly increased from 0 to
1. The rate of cooling (equal to the change in $\lambda$ over one
step) should be much lower than unity in order to ensure that
\scriptsize BAYESYS \normalsize is robust to finding regions of high
posterior value in large prior volumes. Therefore the rate of cooling
is set to a value of 0.1 and multiple runs are performed with
different seed values to check for good convergence. The robustness of
the process is improved by using an ensemble of 10 simultaneous MCMC
chains that communicate after each step and so ensure that they do not
become trapped in local maxima of likelihood.

After burn-in, the program draws samples from the posterior
distribution at $\lambda = 1$, where the total number of samples drawn
is equal to the product of the number of MCMC chains and steps. The
number density of the model samples should then be a good estimate of
the true posterior probability distribution. The resulting output
includes a list of the samples which can be marginalised over to
obtain posterior distributions for the cluster properties, as well as
an estimate of the evidence.

\section{Testing the Model}

We test the ability of the entropy-based model described in
Section\,\ref{section:entropy_model} to reproduce the known properties
of galaxy clusters by fitting to simulated SZ and X-ray surface
brightness data. We simulate SZ data from the CBI2 31\,GHz
interferometer and typical X-ray surface brightness data from the EPIC
camera on \emph{XMM-Newton}.

\subsection{Mock CBI2 SZ data}\label{section:mock_sz_data}

The sky signal for the simulated CBI2 observation is generated from
summing the SZ and intrinsic CMB components. The SZ thermodynamic
temperature for a simulated galaxy cluster is calculated from the
electron pressure using Equation\,\ref{equation:sz_temp}. The CMB
temperature map is constructed from an angular power spectrum
generated using \scriptsize CMBFAST
\normalsize\footnote{http://lambda.gsfc.nasa.gov/toolbox/tb\_cmbfast\_ov.cfm}
\citep{Seljak:1996}, assuming a flat $\Lambda$CDM cosmology with
$\Omega_\rmn{M} = 0.3$, $\Omega_{\Lambda} = 0.7$ and $h = 0.7$. The SZ
and CMB thermodynamic temperature maps are converted to sky brightness
units for each of the 10 frequency channels (26-36\,GHz) using the
differential form of the Planck equation
\begin{equation} {\rmn{d}I_{\nu}\over \rmn{d}T} =
  {2k_\rmn{B}\nu^{2}\over c^{2}}x^{2}\rmn{e}^{x}(\rmn{e}^{x} -
  1)^{-2},
\end{equation}
where $T$ is the thermodynamic temperature and $x =
h\nu/k_\rmn{B}T_\rmn{CMB}$. The two brightness maps are then summed to
form a combined sky simulation.

The sky brightness map is multiplied by the CBI2 primary beam and fast
Fourier transformed to the ($u$,$v$) plane, where the visibility is
then interpolated onto the positions given by a template
observation. The noise weightings associated with each visibility are
then scaled based on the length of the simulated integration. The
effect of the ground spill-over subtraction technique used in real
CBI2 observations \citep[see e.g.][]{Udomprasert:2004} is simulated by
taking the average of the visibilities from two maps with separate
realisations of the CMB, and subtracting this from the mock data
set. This will increase the noise in the data by a factor of
$\sqrt(3/2)$. Once this has been applied, the data are then in the
same form as a real CBI2 data set.

\begin{figure*}
\centering
\includegraphics[width = 0.8\textwidth]{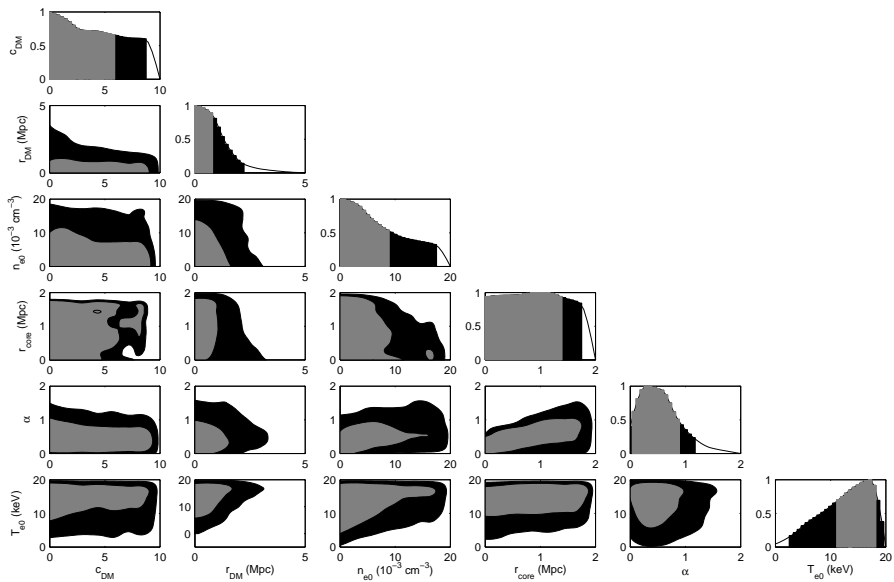}
\caption{The estimated prior probability density for the entropy-based
  model parameters. The constraints arise from the underlying
  assumptions about the cluster physics. The prior ranges encompass
  typical values of these parameters observed to date. The grey scale
  represents estimates of the 68 and 95\,per cent intervals.}
\label{figure:entropy_prior}
\end{figure*}

\subsection{Mock X-ray surface brightness data}\label{section:mock_xray_data}

The X-ray surface brightness is calculated from the known electron
density and temperature distribution of a simulated cluster by using
Equation\,\ref{equation:xray_sb}. The resulting surface brightness map
is then multiplied by the count rate conversion factor of a typical
instrument (the EPIC camera on \emph{XMM-Newton}), which is assumed to
be constant across the 0.5-2\,keV energy band. We use a typical
neutral hydrogen column density of $10^{20}\,\rmn{cm}^{-2}$, and an
integration time equal to 10\,ks. The map is then multiplied by the
solid angle per pixel ($\Delta\Omega$) to convert the data to the
expected counts per pixel. The number of counts on each pixel is then
related to the surface brightness and the pixel solid angle by the
following expression
\begin{equation}\label{equation:no_counts}
  N_\rmn{cnts} =
  \left({S_\rmn{X}\over\rmn{erg}\,\rmn{s}^{-1}\,\rmn{cm}^{-2}\,\rmn{sr^{-1}}}\right)\left(\Delta\Omega\over\rmn{sr}\right)5\times10^{15}\,\rmn{cnts}.
\end{equation}
Noise is introduced into the map by applying the Poisson distribution
to each pixel, with the mean equal to the photon count. For the
purposes of the mock observation we do not take into account the point
spread function, the effect of which is assumed to be insignificant on
the cluster scales that are being modelled. We also do not the take
into account the residual background X-ray emission which would need
to be considered for a real observation.

The X-ray surface brightness profile data are then constructed by
binning the photon counts map into 5\,arcsec radial bins and
converting from counts to physical units. The error in the mean value
associated with each bin is just the square root of the bin value
divided by the number of pixels that contributed to that radial
bin. Hence at moderately large radii the number of pixels contributing
to each bin is large and so the statistics can be approximated by a
Gaussian distribution.

\section{Results}

\subsection{The Prior}

It is important to first understand the intrinsic constraints
introduced by the prior before performing a joint fit to the data. The
prior includes both the volume of the parameter space being considered
and the assumptions upon which the model is
based. Figure\,\ref{figure:entropy_prior} shows the estimated prior
distribution for each of the parameters in the entropy-based
model. This plot was constructed by running the analysis pipeline with
all of the model constraints in place, but with no SZ or X-ray surface
brightness data. The MCMC chains are allowed to search parameter space
within the constraints of the model until a large number of samples
are generated ($\sim 100,000$) and thus produce a reasonable
estimation of the prior distribution.

Constraints on the parameters arise from the assumption that the
intra-cluster medium is in hydrostatic equilibrium and that the
pressure profile must be physical at all radii. The model also prefers
higher central electron temperature values due to the constraint on
the pressure. In order for the condition of hydrostatic equilibrium to
be satisfied, large mass values can only correlate to relatively high
temperatures, and this will provide a constraint on the value of
$r_\rmn{DM}$ which is seen in the figure. The nature of
interferometric data is such that they are not able to constrain a
signal that is non-varying over the angular scales to which the
instrument is sensitive. Therefore in the case of low signal-to-noise
data it would be possible to obtain solutions within the acceptable
prior range of the model, where the temperature increases rapidly at
large radii and while the density profile decreases, leading to an SZ
signal with a large constant additive component. To avoid this clearly
unphysical situation we impose the constraint that the electron
temperature must either be constant or decreasing at radii larger than
the virial radius. The principal effect of this constraint is to
significantly reduce the probability of having an nonphysical high
value for the entropy power law parameter $\alpha$ at large radii.

\subsection{Self-Consistency}

\begin{figure*}
  \centering
  \includegraphics[width=0.8\textwidth]{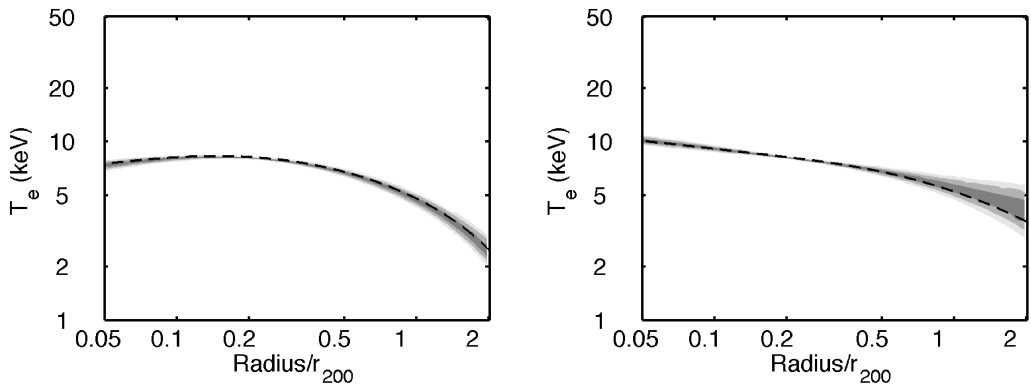}
  \caption{The estimated posterior probability density for the
    temperature as a function of radius, from joint fitting the
    entropy-based model to mock high signal-to-noise CBI2 SZ and X-ray
    surface brightness data, for cool core (\emph{left}) and non-cool
    core (\emph{right}) clusters constructed from the entropy-based
    model. The grey scale represents the 68, 95 and 99\,per cent
    intervals. The dashed lines represent the true simulation values.}
\label{figure:self_consistency_results}
\end{figure*}

We construct mock CBI2 SZ and X-ray surface brightness data from the
entropy-based model in order to test that, given idealised high
signal-to-noise, the correct values for cluster properties can be
returned from model fitting to the data. Two spherically-symmetric
model clusters are constructed for the self-consistency test, one with
a cool-core and the other with a non-cool core. For this test the data
have high signal-to-noise and the intrinsic CMB features are not
included in the SZ signal. The estimated posterior probability density
for the temperature profile of each model cluster is shown in
Figure\,\ref{figure:self_consistency_results}. The results show that
correct unbiased estimates of the cluster properties are reproduced by
fitting the entropy-based model to idealised data constructed from the
same model.

\subsection{Hydrodynamic/N-body Simulations}

\begin{figure*}
\centering
\includegraphics[width=0.8\textwidth]{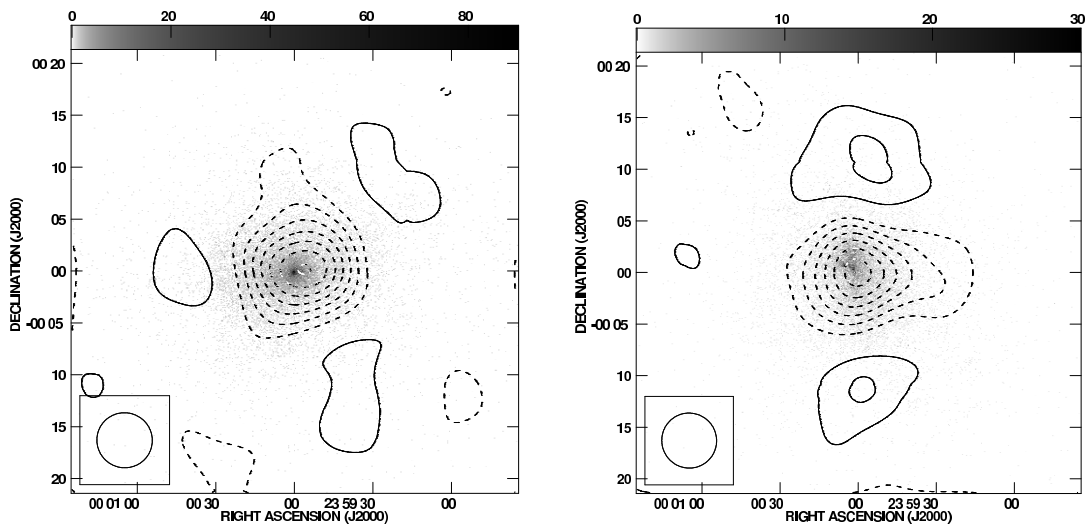}
\caption{Mock observations of the X-ray surface brightness
  (\emph{grey-scale}) and SZ temperature decrement (\emph{contours}),
  for the simulated clusters FB1 (\emph{left}) and FB3 (\emph{right})
  from \citet{Kay:2008}. The grey-scale is in units of counts per
  pixel for a generic X-ray observation. The contours represent
  multiples of the 2\,$\sigma$ noise level from a typical 31\,GHz
  observation using the CBI2 array. The full-width half-maximum of the
  CBI2 synthesised beam is indicated in the bottom left-hand corner of
  each map.}\label{figure:simulation_maps}
\end{figure*}

\begin{figure*}
  \centering
  \includegraphics[width=1.0\textwidth]{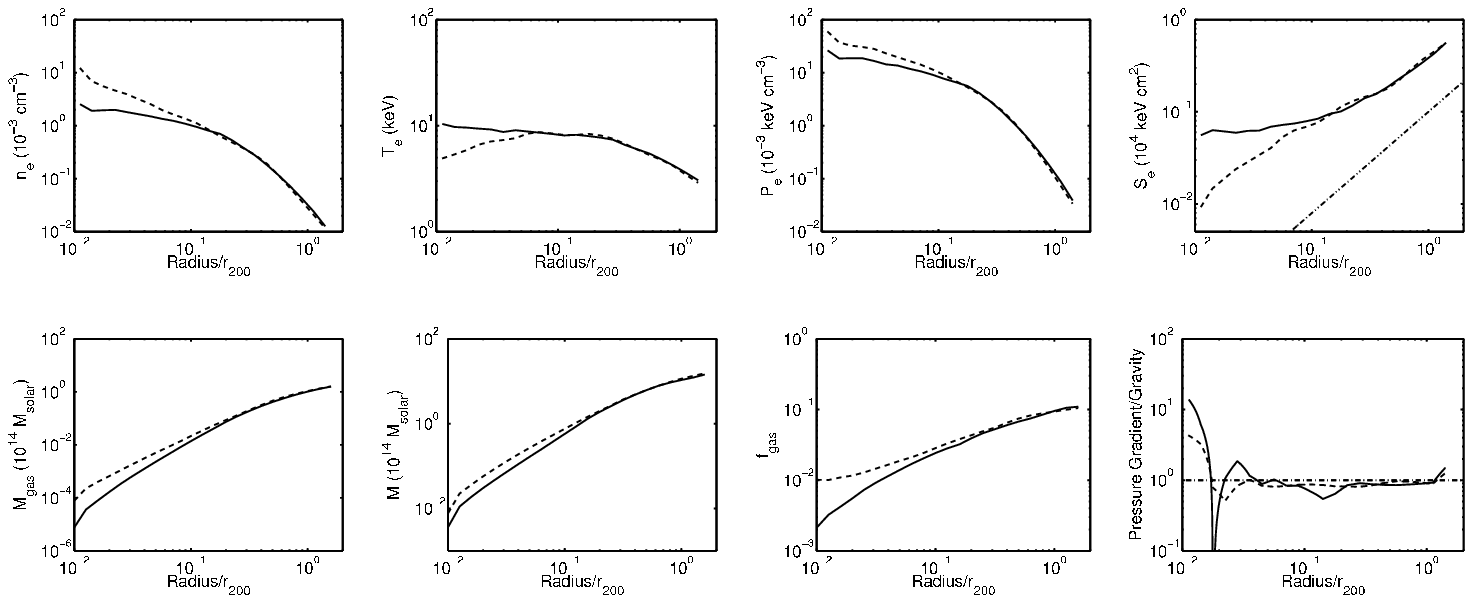}
  \caption{The physical properties of the simulated clusters FB1
    (\emph{dashed line}) and FB3 (\emph{solid line}) as a function of
    radius. The dot-dashed line in the frame showing the radial
    entropy represents the power law predicted by simulations. The
    dot-dashed line in the frame showing the ratio of the pressure
    gradient to gravity represents a cluster in hydrostatic
    equilibrium. FB1 is cool-core relaxed cluster and FB3 is non
    cool-core.}
\label{figure:simulation_profiles}
\end{figure*}

Cosmological $N$-body/hydrodynamic simulations of two clusters from
\cite{Kay:2008} are used to test the entropy-based model. The
simulations were run the with the {\sc Gadget2} code
\citep{Springel:2005} that uses the Particle-Mesh and tree algorithms
to calculate gravitational forces and Smoothed Particle Hydrodynamics
to model the gas. Additionally, the gas was allowed to cool
radiatively, leading to a decrease in entropy, and form collisionless
stars at low temperature ($T<10^{5}$\,K) and high density
($n_\rmn{H}>10^{-3}\rmn{cm^{-3}}$). Thermal energy was also injected
into such regions (feedback) leading to a large increase in the
entropy of the local gas. This phenomenological approach to re-heating
and galaxy formation in the cluster was found by \cite{Kay:2004b} to
reproduce the observed $L_\rmn{X}-T_\rmn{X}$ relation, and so provides
a good test for the entropy-based model. The sub-sample used in this
work consists of two clusters with similar total masses
($\sim10^{15}$\,M$_\odot$) and temperatures ($\sim5$\,keV), but with
quite different merger histories. FB1 has a cool core and is in a
relaxed state, while FB3 is a non-cool core cluster and recently
underwent a strong merger at $z\sim0.4$. Therefore our model should
not only be able to reproduce the global properties of each cluster,
but also correctly constrain the different profiles at large radii.

Maps of $\Delta T_\rmn{SZ}$ (at 31\,GHz) and $S_\rmn{X}$ (in the 0.5 -
2.0\,keV band) were generated for each cluster by calculating the
electron density and temperature properties for every hot particle
within a cylinder of length 6\,$r_{200}$ and projected radius
3\,$r_{200}$. Each particle contribution is smoothed and projected
along the length of the cylinder onto a 1024 $\times$ 1024 2D pixel
array, using the projected version of the \scriptsize GADGET2
\normalsize SPH kernel. The SZ decrement and X-ray surface
brightnesses were calculated using Equations\,\ref{equation:sz_temp}
and \ref{equation:xray_sb}. The CBI2 SZ and X-ray surface brightness
maps for each simulated cluster are shown in
Figure\,\ref{figure:simulation_maps} with realistic noise applied.
Note that, for a fair comparison, the same $\Lambda(T_\rmn{e})$
cooling function is used to generate the simulated X-ray surface
brightness map that is also used when producing the aforementioned
analytical model. The maps are centred about the most
gravitationally-bound dark matter particle and thus the most globally
symmetric point. For the relaxed cluster FB1, this coincides with the
brightest SZ and X-ray surface brightness points. In the case of FB3,
this cluster underwent a recent merger event that has produced an
asymmetric core region, and therefore the brightest pixels in the maps
are offset from the centre. However the symmetry of the global
structure of this cluster is centred about the central pixel of the
map and not the fine substructure of the brightest peaks. This
therefore acts to flatten the central part of the electron density
profile, and hence also the X-ray surface brightness.

The profiles for the physical properties of the simulated clusters are
shown in Figure\,\ref{figure:simulation_profiles}. Clusters of
galaxies typically depart from hydrostatic equilibrium within their
core region due to physical processes other than gravitational
collapse, such as disturbance from recent merger activity, pre-heating
from active galactic nuclei and radiative cooling. Both simulated
clusters exhibit this behaviour within a radius of $\la
100\,\rmn{kpc}$, and their gas properties diverge considerably in this
region. We therefore choose to ignore the central cluster core in
fitting to the data since the high gas density, and hence X-ray
surface brightness, will strongly bias the results.

\subsection{Results from simulated data}\label{section:simulations_results}

\begin{figure*}
\centering
\includegraphics[width = 1.0\textwidth]{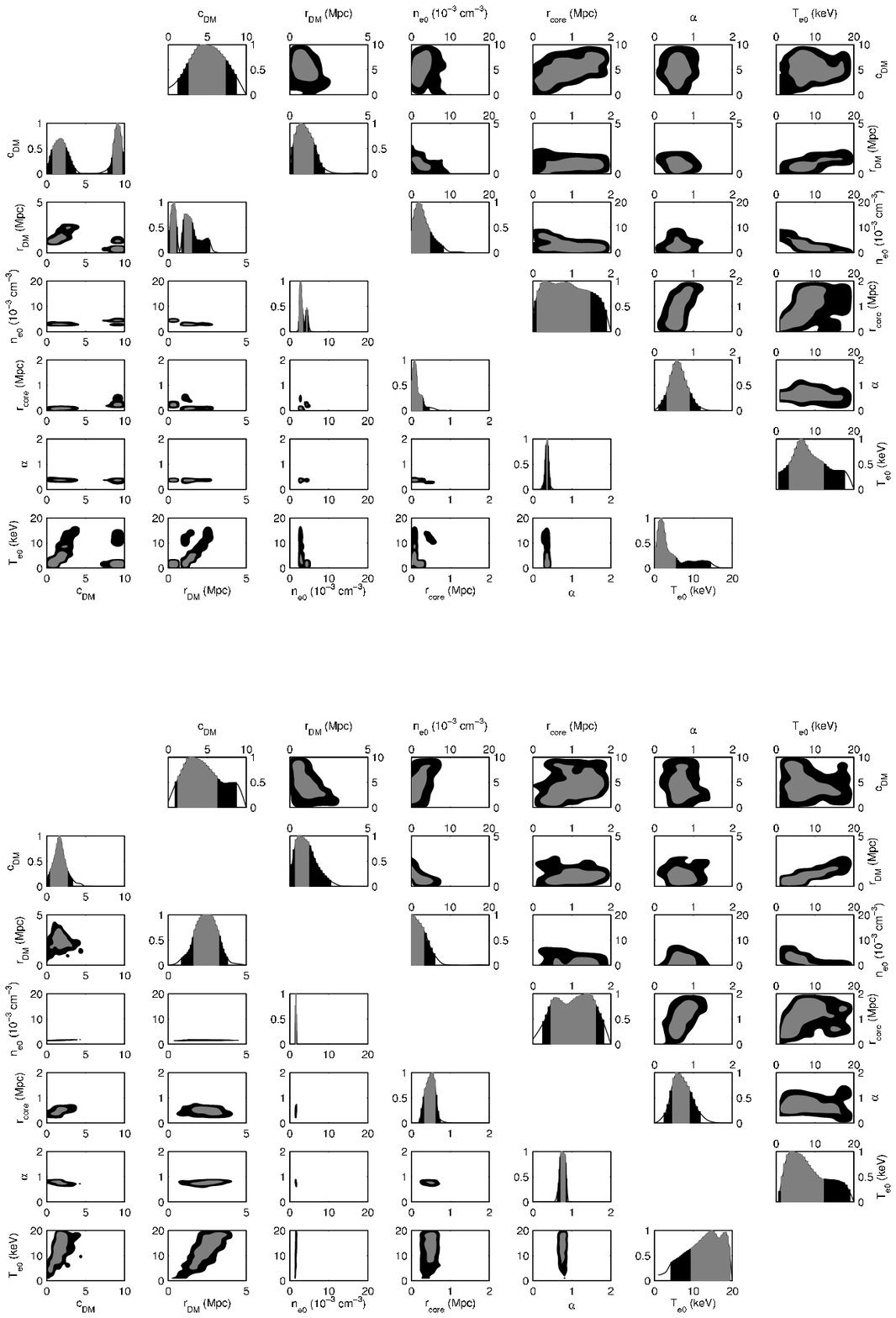}
\caption{The estimated posterior probability density for the
  entropy-based model parameters from separately fitting to simulated
  high signal-to-noise CBI2 SZ data (\emph{top triangle}) and X-ray
  surface brightness data (\emph{bottom triangle}). The top and bottom
  plots represent fits for cluster FB1 and FB3 respectively. The
  grey-scale represents the 68 and 95\,per cent intervals.}
\label{figure:kay_simulations_sz_xrays_only}
\end{figure*}

Mock SZ data and X-ray surface brightness data are constructed from
the simulated cluster maps using the method described in sections
\ref{section:mock_sz_data} and \ref{section:mock_xray_data}. We fit
the entropy-based model to both high and low signal-to-noise simulated
data in order to investigate possible systematics introduced into the
derived cluster properties. For the high signal-to-noise case we fit
to idealised data with very low experimental error, no calibration
errors, and no intrinsic CMB signal in the SZ data. Conversely for low
signal-to-noise data we simulate larger experimental error, and
introduce calibration error and an intrinsic CMB component in the SZ
data. The calibration errors are introduced as nuisance parameters to
be marginalised over and are distributed by a Gaussian prior with
$\sigma_\rmn{cal}$ given by the quoted error value. A calibration
error of 5\,per cent is used for the SZ data, typical of the
calibration of CBI2 visibility data, and the X-ray surface brightness
data typically contain a calibration error of 10\,per cent
\citep{Andersson:2004}.

Figure\,\ref{figure:kay_simulations_sz_xrays_only} shows the
constraints introduced from separately fitting to high signal-to-noise
SZ and X-ray data. The SZ data constrain the integrated line-of-sight
pressure and therefore generate an anti-correlation between the
central electron density and temperature parameters. In addition the
SZ data reduce the likelihood of having large central electron density
and temperature values since these would generate large SZ signals
that are inconsistent with the data. The X-ray surface brightness data
are proportional to the integrated square of the line-of-sight
electron density and therefore strongly constrain the central electron
density parameter. The X-ray data is of relatively higher resolution
than the SZ data and is therefore much more dependent upon the shape
of the entropy profile providing strong constraints on $r_\rmn{core}$
and $\alpha$. The high signal-to-noise surface brightness data do not
provide a strong constraint on the temperature and mass of the
cluster. In the case of FB1 the X-ray surface brightness data appear
to generate a series of high likelihood peaks within parameter
space. However the data are unable to distinguish between these in the
absence of an additional constraint from the SZ data.

\begin{table}
\centering
\begin{tabular}{lcccc}
  \hline
  Name & Data & $Y_\rmn{200}\,D_\rmn{A}^{2}$ & $M_\rmn{200}$ & $f_\rmn{gas,200}$ \\     
  \hline
  FB1 & Simulation & 0.931 & 11.6 & 0.094 \\
\\
  & High S/N & $0.986^{+0.011}_{-0.016}$ & $11.4^{+0.2}_{-0.3}$ & $0.098^{+0.003}_{-0.002}$ \\
\\
  & Low S/N & $1.25^{+0.41}_{-0.133}$ & $15.8^{+3.8}_{-5.1}$ & $0.064^{+0.016}_{-0.009}$ \\
 \\
 FB3 & Simulation & 0.863 & 10.7 & 0.096 \\
\\
  & High S/N & $0.89^{+0.03}_{-0.02}$ & $11.3^{+0.4}_{-0.4}$ & $0.091^{+0.002}_{-0.003}$ \\
\\
  & Low S/N & $1.05^{+0.40}_{-0.23}$ &$15.8^{+5.0}_{-5.4}$ & $0.064^{+0.020}_{-0.020}$\\
  \hline
\end{tabular}
\caption{Results from joint entropy-based model fits to the simulated clusters. Column 1 gives the cluster name; column 2 the data description; column 3 the integrated comptonisation within the virial radius, in units of $10^{-4}\,\rmn{Mpc}^{2}$; column 4 the total mass within the virial radius, in units of $10^{14}\,M_{\odot}$, column 5 the gas mass fraction within the virial radius. Errors represent the 68\,per cent interval.}
\label{table:simulation_properties}
\end{table}

\begin{figure*}
\centering
\includegraphics[width = 1.0\textwidth]{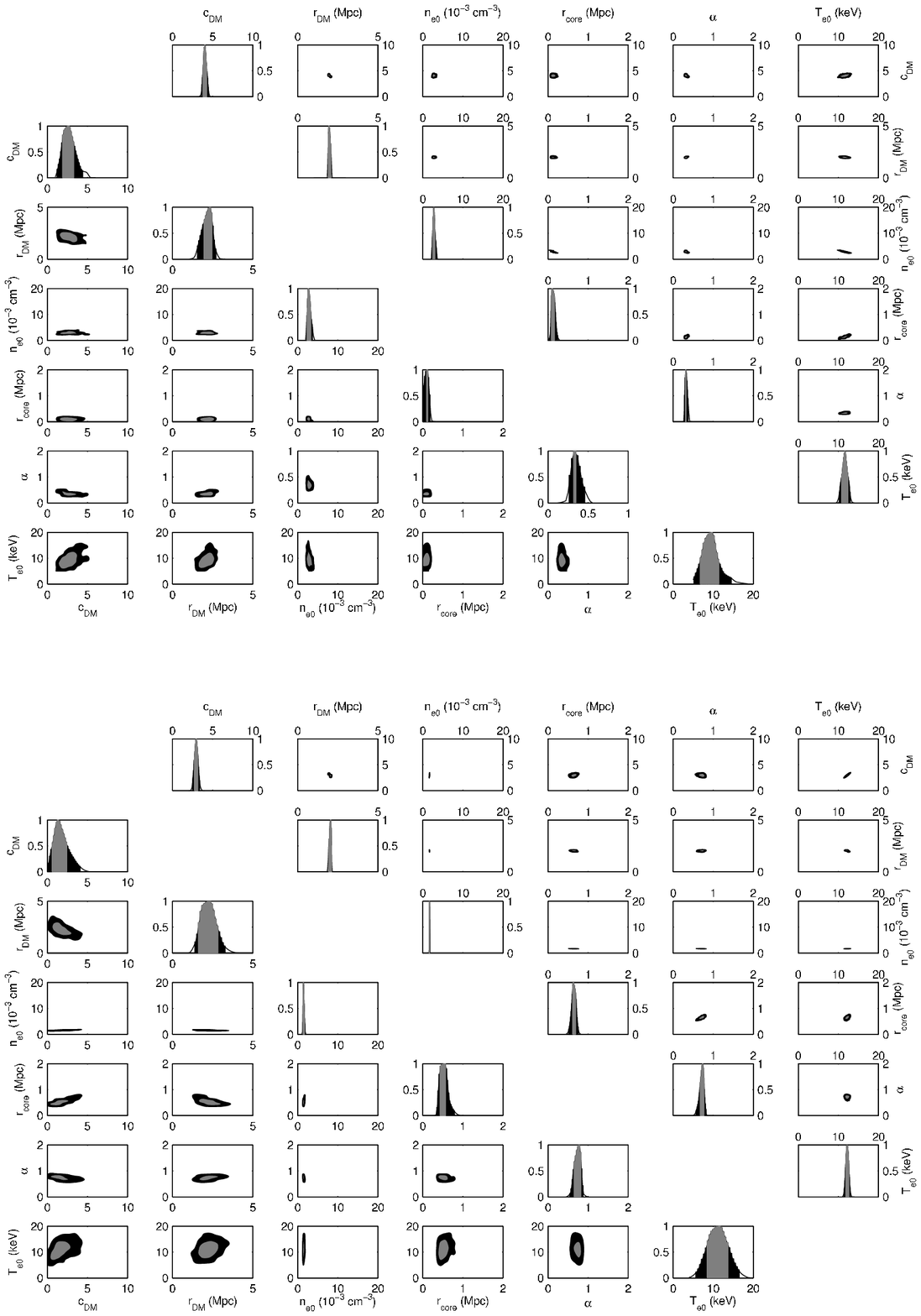}
\caption{The estimated posterior probability density for the
  entropy-based model parameters from joint fitting to simulated high
  signal-to-noise (\emph{top triangle}) and low signal-to-noise
  (\emph{bottom triangle}) CBI2 SZ and X-ray data. The top and bottom
  plots represent fits for clusters FB1 and FB3 respectively. The
  grey-scale represents the 68 and 95\,per cent intervals.}
\label{figure:kay_simulations_entropy_params}
\end{figure*}

\begin{figure*}
\centering
\includegraphics[width = 1.0\textwidth]{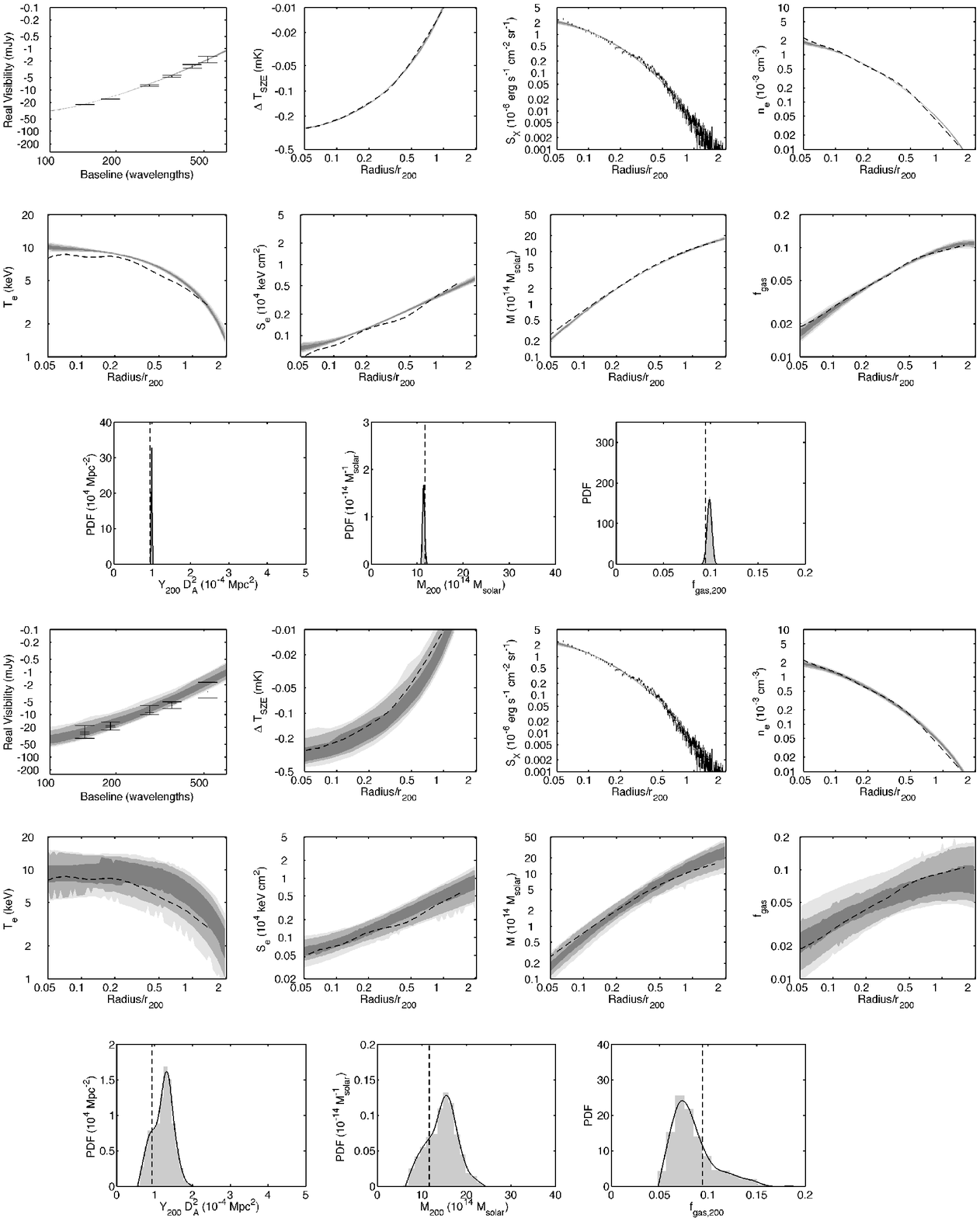}
\caption{The estimated posterior probability density for properties
  of Cluster FB1 from joint fitting the entropy-based model to simulated
  high signal-to-noise (\emph{top}) and low signal-to-noise
  (\emph{bottom}) CBI2 SZ and X-ray data for cluster FB1. The grey
  scale represents the 68, 95 and 99\,per cent intervals. The dashed
  lines represent the true simulation values and the error-bars the
  mock data.}
\label{figure:kay_simulations_entropy_results}
\end{figure*}

\begin{figure*}
\centering
\includegraphics[width = 1.0\textwidth]{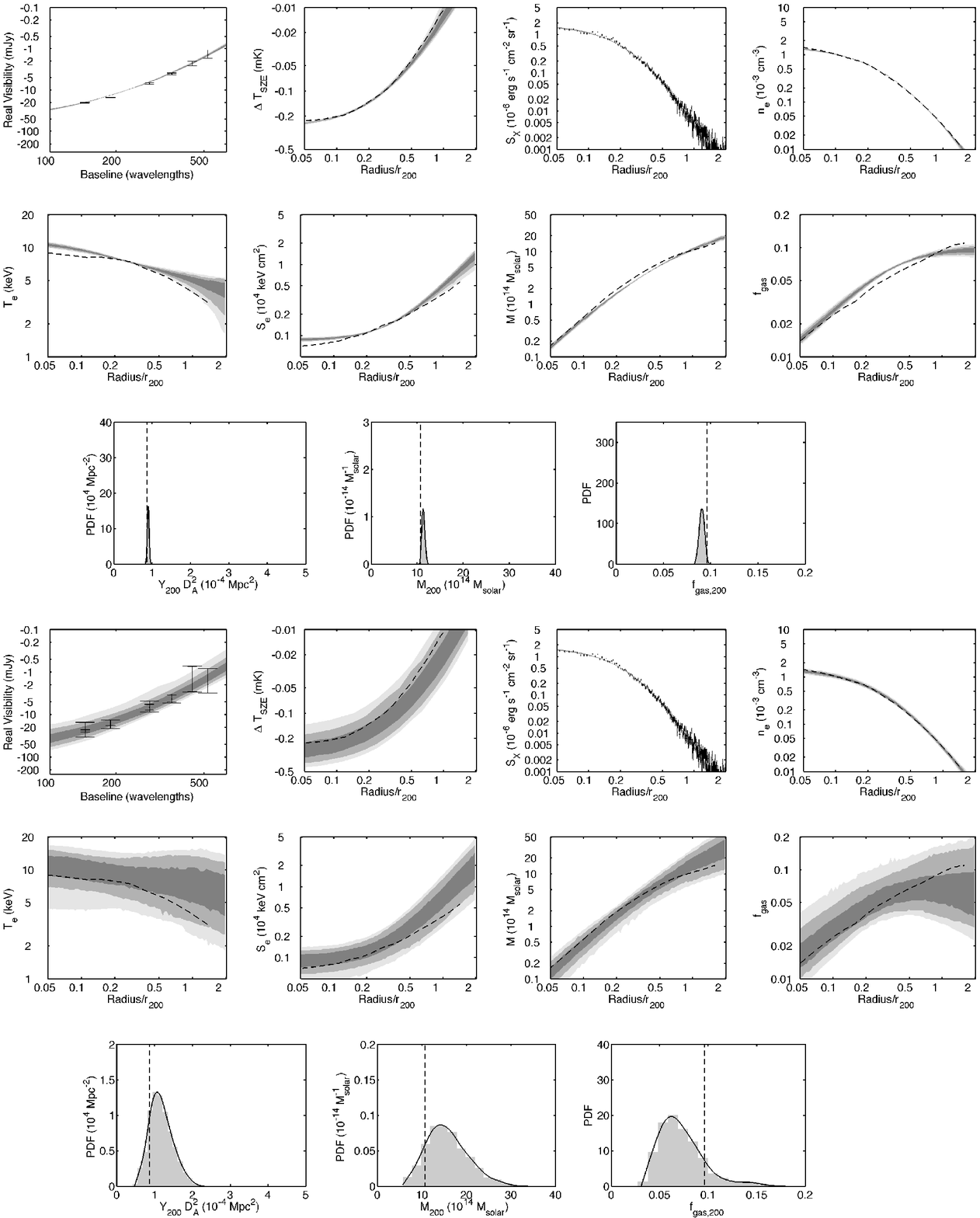}
\contcaption{Cluster FB3.}
\end{figure*}

Results from fitting simultaneously to both data sets, for both the
high and low signal-to-noise cases, are shown in
Table\,\ref{table:simulation_properties} and Figures
\ref{figure:kay_simulations_entropy_params} and
\ref{figure:kay_simulations_entropy_results}, where estimates of the
posterior distribution for the model parameters, profiles of the
cluster properties and selected enclosed quantities are plotted. 

The model parameters are constrained by the combination of both data
sets. For the purposes of comparison with the known simulated cluster
properties the radial profiles are scaled by the known true $r_{200}$
value, and all global quantities are calculated by integrating within
this radius. In the high signal-to-noise regime the radial profiles
and global parameters of the true physical properties are reproduced
well by the parametric fit, with systematic error $< 10$\,per cent at
$r_{200}$. This result is fairly consistent with \cite{Kay:2004b}, who
find that the hydrostatic mass estimated from a combination of $\beta$
model fits to the X-ray surface brightness data and spatial
temperature information agrees to within $10$\,per cent ($r \la
r_{500}$) for the simulated clusters which contain feedback. They find
that the presence of feedback produces a higher degree of
thermalisation than if the cluster were simply described by a
non-radiative model and so the estimated mass is close to the true
value. In the low signal-to-noise regime the increase in noise leads
to much larger widths in the probability distributions of the cluster
quantities and, while shifting the peaks of the distributions, are
still consistent with the known simulation values within the errors.

Table\,\ref{table:evidence_values} gives the natural logarithm of the
evidence for different models given the data. The evidence values for
both low and high signal-to-noise regimes support the inclusion of
hyper-parameters. The ratios of best fit values for the SZ and X-ray
hyper-parameters in each regime are 15:1 and 3.5:1 (high
signal-to-noise), and 2.5:1 and 1.1:1 (low signal-to-noise), for FB1
and FB3 respectively, indicating that the hyper-parameters weight the
likelihood in favour of the SZ data. While the CBI2 SZ data are
insensitive to variation on scales smaller than 500\,kpc, the X-ray
data are much more sensitive to variations over this range. The nature
of the parametric model is such that it is inherently smooth on these
smaller scales; hence clumping of the gas will deviate the model from
the X-ray data, while still producing a relatively good fit to the SZ
data. Therefore the weights will typically favour the SZ data relative
to the X-ray surface brightness. In the case of the FB3 data, the
relative weightings are more similar since the X-ray data are a closer
match to the model over the considered radius range. A reduction in
signal-to-noise leads to a decrease in the ratios of hyper-parameters
for each cluster, and so both the SZ and X-ray data sets provide a
more consistent match to the model.

\subsection{Comparison with the isothermal $\beta$ model}\label{section:isobeta_results}

\begin{figure*}
\centering
\includegraphics[width = 1.0\textwidth]{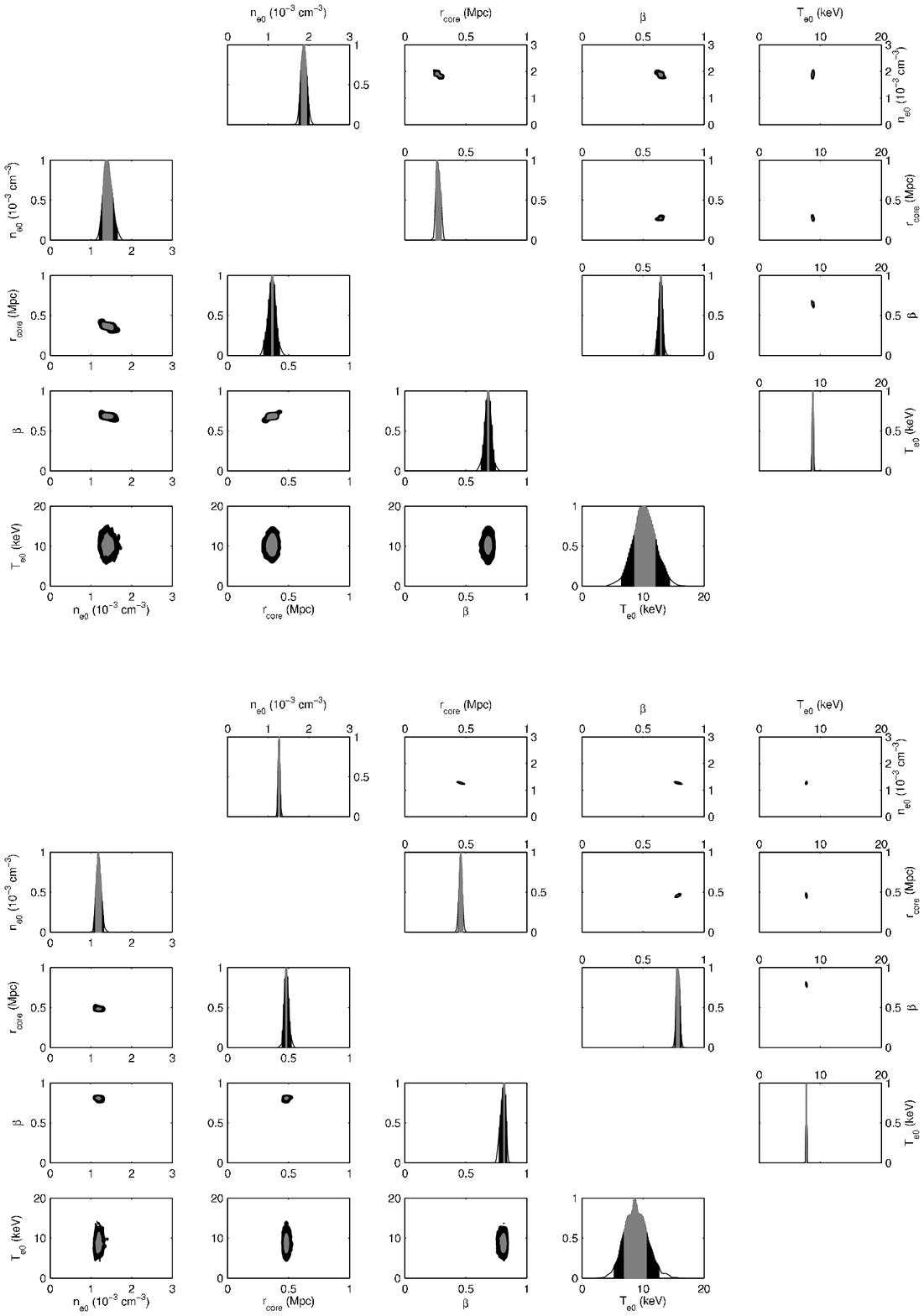}
\caption{The estimated posterior probability density for the
  isothermal $\beta$ model parameters from joint fitting to simulated
  high signal-to-noise (\emph{top triangle}) and low signal-to-noise
  (\emph{bottom triangle}) CBI2 SZ and X-ray data. The top and bottom
  plots represent fits for clusters FB1 and FB3 respectively. The
  grey-scale represents the 68 and 95\,per cent intervals.}
\label{figure:kay_simulations_isothermal_beta_params}
\end{figure*}

\begin{figure*}
\centering
\includegraphics[width = 1.0\textwidth]{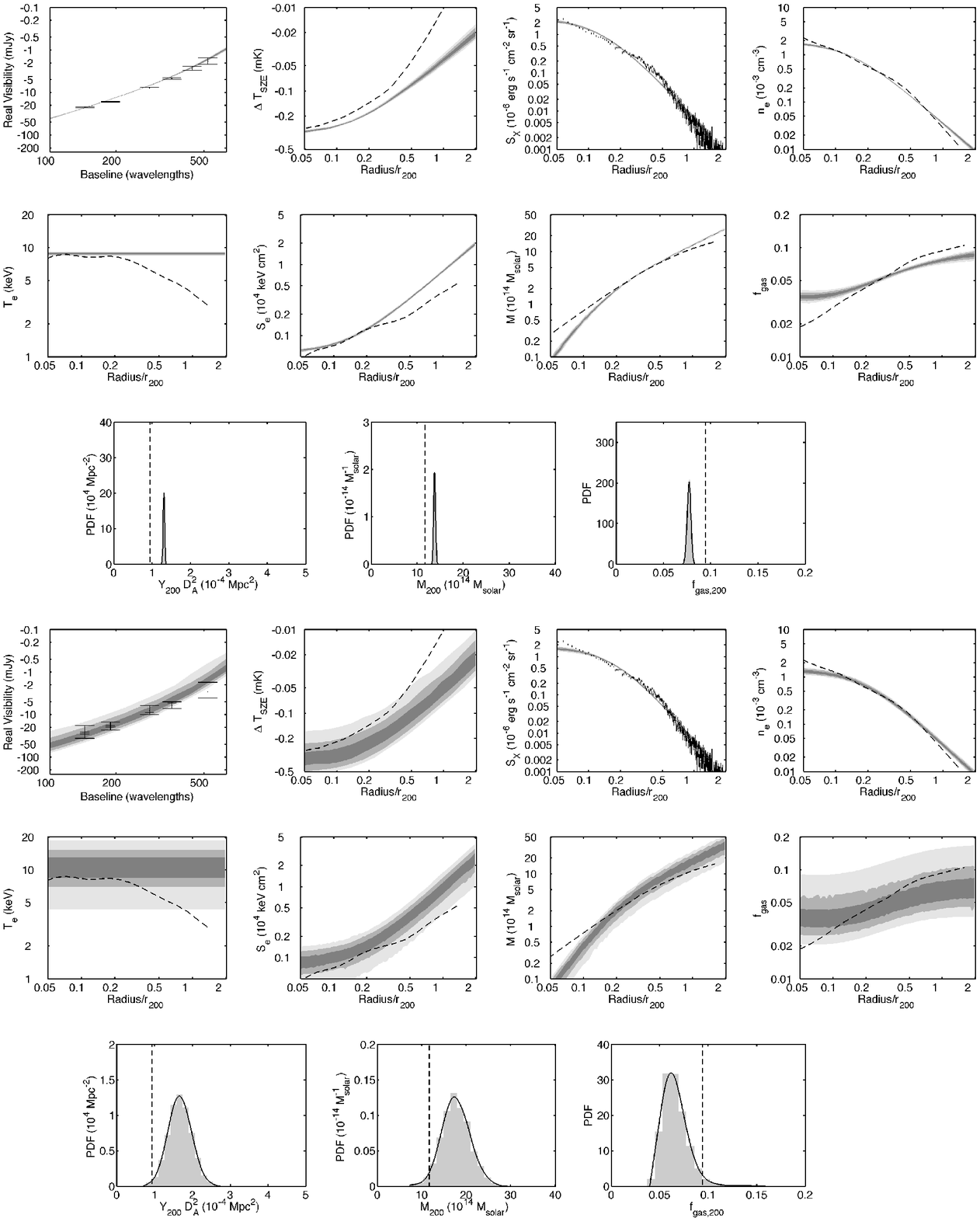}
\caption{The estimated posterior probability density for properties of
  Cluster FB1 from joint fitting the isothermal $\beta$ model to
  simulated high signal-to-noise (\emph{top}) and low-signal-to-noise
  (\emph{bottom}) CBI2 SZ and X-ray data for cluster FB1. The grey
  scale represents the 68, 95 and 99\,per cent intervals. The dashed
  lines represent the true simulation values and the error-bars the
  mock data.}
\label{figure:kay_simulations_isothermal_beta_results}
\end{figure*}

\begin{figure*}
\centering
\includegraphics[width = 1.0\textwidth]{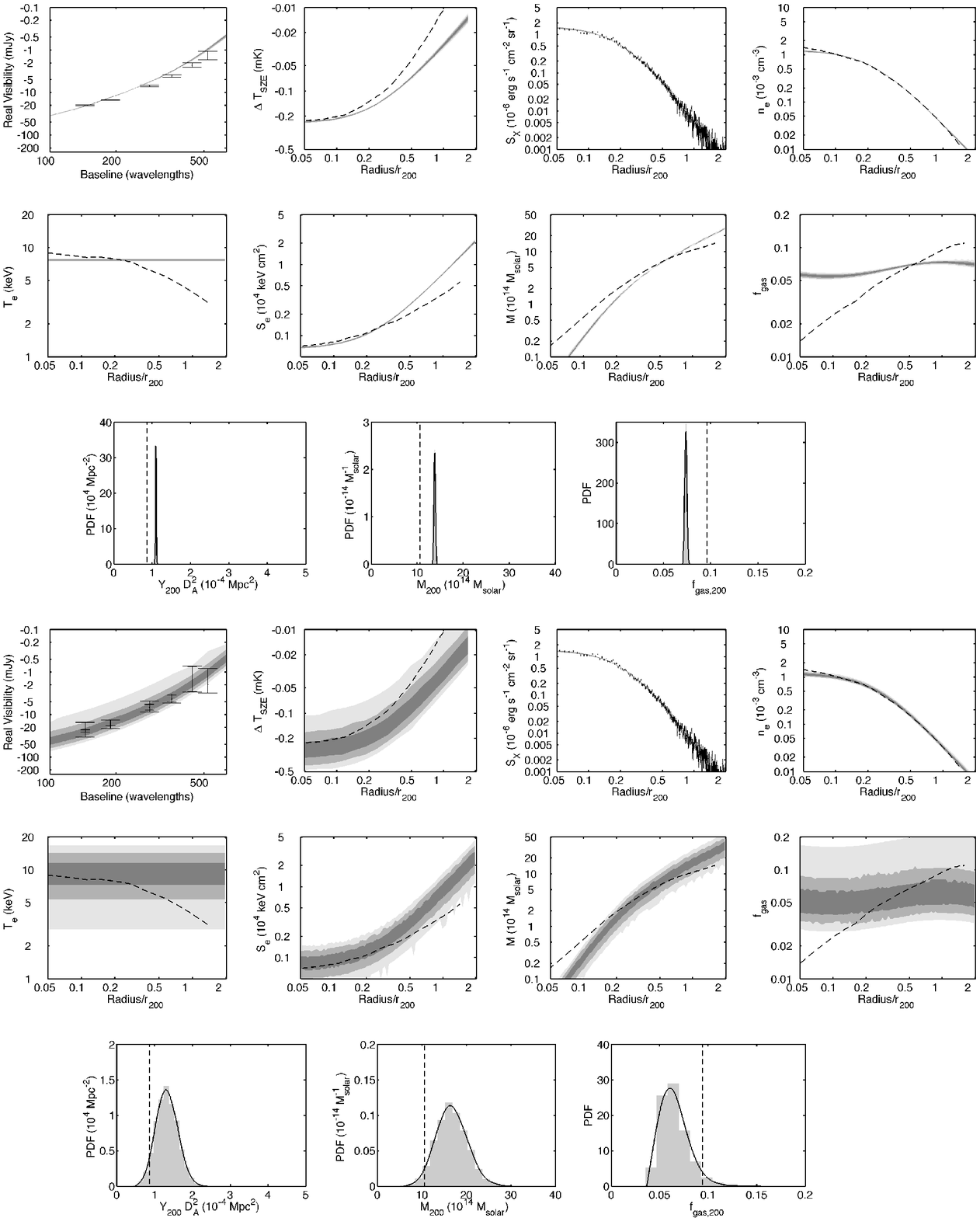}
\contcaption{Cluster FB3.}
\end{figure*}

Even though the simulated clusters are clearly not isothermal, for the
purposes of comparison to previous work and to demonstrate the
systematic differences in derived cluster properties using the two
different models, we perform joint fits using the single isothermal
$\beta$ model. The electron density and temperature are then given by
\begin{eqnarray}
n_\rmn{e} & = & n_\rmn{e0}\left(1 + \left({r\over r_\rmn{core}}\right)^{2}\right)^{-3\beta/2} \\
T_\rmn{e} & = & T_\rmn{e0},
\end{eqnarray}
where $n_\rmn{e0}$ and $T_\rmn{e0}$ are the central electron density
and temperature respectively, $r_\rmn{core}$ is the core radius, and
$\beta$ is a parameter that determines the large scale behaviour of
the electron density.

Figures \ref{figure:kay_simulations_isothermal_beta_params} and
\ref{figure:kay_simulations_isothermal_beta_results} show the
resulting posterior distributions for the model parameters, the
profiles, and global values within $r_{200}$. There is a strong
systematic difference between the estimated and true simulation values
when using this model, resulting from over-estimates of the electron
temperature and SZ decrement. The result is a systematic over-estimate
of the total mass of $\sim$ 20\,per cent at $r_{200}$ and an
under-estimate of the total mass at $r < r_{500}$. These results are
consistent with the findings by \cite{Kay:2004b} who measure similar
errors in estimating the total mass based upon the isothermal
model. The systematic error in derived cluster properties is seen in
both the high and low signal-to-noise scenarios, and therefore the
introduction of both thermal noise and intrinsic CMB anisotropy is not
enough to dominate over the effects of the intrinsic isothermal
$\beta$ model discrepancy with the simulations.

It is informative to compare the relative quality of the fit that the
entropy-based and single isothermal $\beta$ models give, based upon
their respective logarithmic evidence values. In almost all the cases
the evidence for the isothermal $\beta$ model is significantly lower
than that of the entropy-based model, with the exception of the low
signal-to-noise scenario for the FB3 simulated cluster. This is due to
the flat core nature of the X-ray surface brightness profile for this
cluster, as a result of the displacement of the gas peak in the
central region. As such when significantly lower signal-to-noise data
is used the isothermal $\beta$ model, with its flat core behaviour at
small radii, provides an equally good fit (if not slightly better)
than the entropy-based model. However in the case of the FB1 simulated
cluster, the X-ray profile is strongly peaked in the central region
and therefore the isothermal $\beta$ model provides a significantly
poorer fit to both high and low signal-to-noise data. This can be seen
upon visual inspection of the electron density profiles for FB1 in
Figure\,\ref{figure:kay_simulations_isothermal_beta_results}, where
the isothermal $\beta$ model fails to provide a suitable fit at both
small and large radii.

The evidence values in Table\,\ref{table:evidence_values} support the
inclusion of hyper-parameters in both the high and low signal-to-noise
scenarios for the isothermal $\beta$ model. The relative weightings of
each the SZ and X-ray data sets are to 30:1 (FB1) and 3:1 (FB3) for
the high signal-to-noise data, and 4:1 and 1.1:1 for the corresponding
low signal-to-noise data. The likelihood calculation is therefore
weighted in favour of the SZ data, except in the case of the low
signal-to-noise FB3 data where both are almost equally favoured.

\begin{table}
\centering
\begin{tabular}{lccccc}
  \hline
  &  & \multicolumn{2}{c}{High SNR} & \multicolumn{2}{c}{Low SNR} \\     
  Name & Model & $\mathcal{M}_{0}$ & $\mathcal{M}_{1}$ & $\mathcal{M}_{0}$ & $\mathcal{M}_{1}$ \\
  \hline
  FB1 & Entropy  & 6113 & 6476 &  4757 & 4764 \\
  & Isothermal $\beta$ & 5946 & 6405 & 4559 & 4724 \\
  FB3 & Entropy & 6147 & 6273 & 4807 & 4816 \\
  & Isothermal $\beta$ & 6131 & 6231 & 4810 & 4817 \\
  \hline
\end{tabular}
\caption{The natural logarithm of the evidence (the probability of the
  data given a particular parametric model, see
  Equation\,\ref{equation:evidence}) for high and low signal-to-noise
  CBI2 SZ and X-ray surface brightness data. Note that the absolute
  value of the evidence is dependent upon the data, and so the
  quantity of interest is the difference between these values. Models
  that include SZ and X-ray hyper-parameters are represented by
  $\mathcal{M}_{1}$ and those that do not are represented by
  $\mathcal{M}_{0}$.}
\label{table:evidence_values}
\end{table}

\section{Discussion and Conclusions}

We have developed a parametric model for the gas in galaxy clusters,
based on three physical assumptions: the dark matter follows an NFW
profile, the gas entropy can be described by a power law with a
flattened core, and the gas is in hydrostatic equilibrium. Using
entropy rather than, for example, the gas density as the basic
parametrization is motivated by the theoretical and observed
self-similarity of entropy profiles in cluster samples.  This model
can be constrained by SZ and X-ray data to give fitted gas and total
matter properties of the cluster. The model has sensible convergence
properties and can be used out to the virial radius of the cluster. By
construction, the model does not allow unphysical or inconsistent
properties of the cluster gas as can happen if, for example, a
parametric fit to the gas density is combined with an unrelated
parametric fit to the temperature.

We have tested the model using two detailed N-body plus hydrodynamical
simulations of massive clusters with contrasting merger histories. In
both cases, using realistic mock data from presently available X-ray
and SZ telescopes, the model is able to accurately fit both the
integrated cluster parameters and their radial profiles. If
high-quality data with very low noise are simulated, the cluster
parameters are returned with essentially no bias. Our fitting code
includes both random noise and systematic calibration errors in the
data and fully includes the effect of contamination from primordial
CMB fluctuations and radio sources in the SZ data. We also use a
hyper-parameter approach to scale the relative constraints from the
two data sets. Comparison with the widely-used isothermal $\beta$
model confirms previous results that this model can result in
significant biases in fitted cluster parameters
\citep[e.g.][]{Kay:2004b, Hallman:2007}. The quality of the available
SZ data is now high enough to require a more sophisticated modelling
approach, especially with data that are sensitive to the outskirts of
cluster.

This model however remains simplistic in several potentially important
ways. The assumption of hydrostatic equilibrium is clearly broken
badly in the central cores of clusters, and we are forced to ignore
the data in this region. Hydrostatic equilibrium will also be broken
in the main body of the cluster due to bulk motions and other
non-thermal support, although in our simulations this does not seem to
be a significant impediment to measuring accurate cluster profiles. We
do not currently treat the boundary of the cluster in a fully
consistent way -- at some radius the virialised gas must meet a
boundary shock of in-falling material and we do not model the
corresponding step in pressure. We also do not yet consider additional
observation constraints such as X-ray spectral and optical weak
lensing measurements, although these are in principle straightforward
to incorporate in to our analysis framework.

In subsequent papers we will use this model to analyse SZ data from the
CBI2 experiment jointly with relevant X-ray imaging data. 

\section*{Acknowledgments} 

We thank Jamie Leech and Paul Grimes for support with computing
problems and the Oxford E-Science Research Centre for providing
much-needed computing power. We thank Jon Sievers and Steve Myers for
many useful conversations and support of {\sc MPIGRIDDR}. We also
thank Filipe Abdalla and Devinder Sivia for useful conversations on
MCMC methods. JRA acknowledges support from a studentship from the
Science and Technology Facilities Council; ACT acknowledges support
from a Royal Society Dorothy Hodgkin Fellowship.

\bibliographystyle{mn2e}
\bibliography{bibliography}

\bsp

\label{lastpage}

\end{document}